\documentclass[aps,twocolumn,groupedaddress,showpacs]{revtex4}
\usepackage{amsmath}
\usepackage{amsfonts}
\usepackage{amssymb}
\usepackage{graphicx}

\input{tcilatex}

\begin{document}

\title{On open string in generic background}
\author{Liu Zhao}
\affiliation{Institute of Modern Physics, Northwest University, Xian 710069, China}
\author{Wenli He}
\affiliation{Institute of Modern Physics, Northwest University, Xian 710069, China}

\begin{abstract}
A set of consistent Poisson brackets for an open string in generic spacetime
background and NS-NS $B$-field is constructed. Upon quantization, this set
of Poisson brackets lead to spacial \emph{commutative} $D$-branes at the
string ends, showing that noncommutativity between spacial coordinates on
the $D$-branes can be avoided.
\end{abstract}

\maketitle


%Optional argument for running titles on pages
%\title[]{}

%repeat the \author .. \affiliation  etc. as needed
%\email, \thanks, \homepage, \altaffiliation all apply to the current
%author. Explanatory text should go in the []'s, actual e-mail
%address or url should go in the {}'s for \email and \homepage.
%Please use the appropriate macro for the type of information

%\affiliation command applies to all authors since the last
%\affiliation command. The \affiliation command should follow the
%other information
%\affiliation can be followed by \email, \homepage, \thanks as well.

%\hspace{1cm}
%\email{lzhao@nwu.edu.cn}

%\email{hewenli@phy.nwu.edu.cn}

%\date{\today}
%\date{}


%insert suggested PACS numbers in braces on next line

%insert suggested keywords - APS authors don't need to do this


\section{Introduction}

Field theories defined on noncommutative spaces have attracted
considerable attention during the last few years \cite{NC}. This
is largely due to the
study of open string theory in background NS-NS $B$-field. For constant $B$%
-field, a number of papers \cite{SJ,Chu1,Chu2,Arfaei,Braga,Deriglazov1}
%\cite{SJ} \cite{Chu1} \cite{Chu2} \cite{Arfaei} \cite{Braga}
%\cite{Maeno} \cite{Deriglazov1}
by different authors have revealed that the quantization procedure would
lead to noncommutativity at the boundary, i.e. D-branes in the perturbative
string picture, though there are some discrepancies \cite{He} on the way in
which coordinates on the D-branes noncommute.

For string theory to be a fundamental theory, it is reasonable to require
that the description of string theory should be background independent. Of
course the background independence cannot be realized in a verbose way, but
there is a large library of string dualities to which we can resort to
realize in a lose sense the background independence. Moreover, in the study
of noncommutative gauge theories, Seiberg and Witten \cite{SW} found some
famous maps between 1) a noncommutative field theory with a small
noncommutative parameter and a commutative one and 2) one noncommutative
field theory with another with different noncommutative parameters. However,
there are some ambiguities in the Seiberg-Witten maps, some of which can be
resolved \cite{SWAM} by field-dependent gauge transformations and field
redefinitions, whilst others cannot \cite{AMB}.

In an earlier paper \cite{He}, we studied the problem of quantization of
open string in flat spacetime with constant background $B$-field, with the
aim to resolve the discrepancies between different works. The approach we
used is a direct modification of the Poisson brackets at the boundaries,
following the principle of locality \cite{Zhao}. To our astonishment, the
equations determining the modifications are under determined ($2$ equations
for $3$ unknowns), and we found an infinite number of consistent Poisson
structures, each leads to a different quantization scheme. In particular,
all previous results on the same problem are special cases of our result,
and the discrepancies among them are just a choice of Poisson structure from
our result. What is more remarkable is that, in our result, there is a
particular Poisson structure which does not lead to noncommutativity at the
boundaries upon quantization. This means that in the presence of constant
background $B$-field, open string can be quantized \emph{without}
introducing noncommutativity on the D-branes.

In this Letter, we shall consider the Poisson structure for open string in
generic spacetime metric and background $B$-field using the same method as
in our previous paper \cite{He}. Unlike the case with constant $B$-field,
the equations determining the modified Poisson brackets are over determined
and no general solution to them is known to exist. However, there is a
simple solution which, upon quantization, gives rise to \emph{vanishing}
commutators between coordinates at the open string boundaries. In other
words, the $D$-branes corresponding to our solution are still commutative,
in spite of the presence of the generic background $B$-field. It should be
remarked that the same problem has already been studied by Ho and Yeh in
\cite{Ho1}, which contains very different result. However, the result of
\cite{Ho1} is based on the assumption that the spacetime coordinates $X^{i}$
depend only linearly on the world sheet coordinate $\sigma $, i.e. $%
X^{i}=x^{i}+p^{i}\sigma $, which is not valid in general, and it is not
known whether the Poisson brackets (and consequently the commutators after
quantization) obtained there obey Jacobi identities or not.

\section{Poisson structure for open string in generic background}

Now let us consider the problem in detail. The bosonic part of the world
sheet action for open string theory reads,%
\begin{align*}
S & =\frac{1}{4\pi\alpha^{\prime}}\int d^{2}\sigma\sqrt{-g}%
\,[g^{ab}G_{ij}(X)\partial_{a}X^{i}\partial_{b}X^{j} \\
& +2\pi\alpha^{\prime}B_{ij}(X)\varepsilon^{ab}\partial_{a}X^{i}\partial
_{b}X^{j}+\alpha^{\prime}\Phi(X)R^{(2)}],
\end{align*}
where $G_{ij}(X)$, $B_{ij}(X)$ and $\Phi(X)$ are, respectively, the
background metric, antisymmetric tensor and dilaton fields, which are the
only fields in the NS-NS sector. The world sheet metric $(g^{ab})$ is chosen
to have the signature $(1,-1)$, and $\epsilon^{01}=1=-\epsilon^{10}$. It is
commonly known that a quantum string theory cannot exist under arbitrary
background fields $G_{ij}(X)$, $B_{ij}(X)$ and $\Phi(X)$. To ensure quantum
consistency, only those background fields which have quantum Weyl invariance
are allowed. This requirement induces certain constraints on the choice of
the background fields, which, to first order in $\alpha^{\prime}$, are given
by the vanishing of the beta functionals. A crucial observation is that, the
vanishing conditions for the beta functionals are differential equations in
spacetime only which do not involve any world sheet derivatives. Therefore,
we assume throughout this Letter that the background fields $G_{ij}(X)$, $%
B_{ij}(X)$ and $\Phi(X)$ depend only on the spacetime coordinates $X^{i}$
and subject to the vanishing conditions for the beta functionals. Such
background fields are referred to as \emph{generic backgrounds} as compared
to the constant backgrounds. We assume also that the field $B_{ij}$ is
spacetime-filling and invertible, otherwise we shall be considering only the
subset of the spacetime in which $B_{ij}$ is invertible.

Since the background fields we have chosen to study preserve conformal
symmetry, we may use this symmetry to choose a specific gauge for the world
sheet metric, i.e. the flat gauge under which the two dimensional Ricci
scalar $R^{(2)}$ vanishes and the metric $g^{ab}$ reads $(g^{ab})=diag(1,-1)$%
. Under this choice, the action may be rewritten as
\begin{align}
S & =\frac{1}{4\pi\alpha^{\prime}}\int d^{2}\sigma\lbrack
g^{ab}G_{ij}(X)\partial_{a}X^{i}\partial_{b}X^{j}  \notag \\
& \hspace{5mm}+2\pi\alpha^{\prime}B_{ij}(X)\varepsilon^{ab}\partial_{a}X^{i}%
\partial_{b}X^{j}],   \label{3.1}
\end{align}
in which the dilation field $\Phi(X)$ simply decouples.

The variation of (\ref{3.1}) yields the equations of motion
\begin{align*}
&
\partial^{a}\partial_{a}X^{k}+\Gamma_{ij}^{k}(X)g^{ab}\partial_{a}X^{i}%
\partial_{b}X^{j} \\
& \hspace{5mm}-\pi\alpha^{\prime}G^{kl}(X)B_{(ij,\,l)}(X)\epsilon
^{ab}\partial_{a}X^{i}\partial_{b}X^{j}=0
\end{align*}
and the mixed boundary conditions
\begin{equation}
\left. \left[ G_{ik}(X)\partial_{\sigma}X^{i}+2\pi\alpha^{\prime}B_{kj}(X)%
\partial_{\tau}X^{j}\right] \right\vert _{\sigma=0,\pi}=0,   \label{3.3}
\end{equation}
where
\begin{equation*}
\Gamma_{ij}^{k}(X)=\frac{1}{2}G^{kl}\left( \frac{\delta G_{lj}}{\delta X^{i}}%
+\frac{\delta G_{il}}{\delta X^{j}}-\frac{\delta G_{ij}}{\delta X^{l}}%
\right)
\end{equation*}
is the spacetime connection and
\begin{equation*}
B_{(ij,\,l)}(X)=\frac{\delta B_{ij}}{\delta X^{l}}+\frac{\delta B_{jl}}{%
\delta X^{i}}+\frac{\delta B_{li}}{\delta X^{j}}
\end{equation*}
is the cyclic spacetime derivative (the $3$-form field strength) of the
background field $B$.

The canonical conjugate momenta are defined in the usual way,
\begin{equation*}
P_{i}\equiv\frac{\delta\mathcal{L}}{\delta\partial_{\tau}X^{i}}=\frac{1}{%
2\pi\alpha^{\prime}}\left[ G_{ij}(X)\partial_{\tau}X^{j}+2\pi\alpha^{\prime
}B_{ij}(X)\partial_{\sigma}X^{j}\right] ,
\end{equation*}
so that $\partial_{\tau}X^{k}$ can be expressed in terms of $P_{i}$ and $%
X^{j}$ as
\begin{equation}
\partial_{\tau}X^{k}=2\pi\alpha^{\prime}G^{ki}(X)\left[ P_{i}-B_{ij}(X)%
\partial_{\sigma}X^{j}\right] .   \label{3.4}
\end{equation}

In the absence of boundary conditions, the naive Poisson brackets hold for
the world sheet string theory, i.e.
\begin{align}
\left\{ X^{i}(\sigma),X^{j}(\sigma^{\prime})\right\} & =\left\{
P_{i}(\sigma),P_{j}(\sigma^{\prime})\right\} =0,  \label{3.5} \\
\left\{ X^{i}(\sigma),P_{j}(\sigma^{\prime})\right\} &
=\delta^{i}{}_{j}\delta(\sigma-\sigma^{\prime}).   \label{3.6}
\end{align}
However, the appearance of boundary conditions (\ref{3.3}) makes the naive
Poisson structure (\ref{3.5})-(\ref{3.6}) inconsistent at the end points of
the open string, and one has to modify (\ref{3.5}) and (\ref{3.6}) to get a
consistent Hamiltonian formalism of the world sheet theory.

Since the inconsistency only appears at the string end points, one needs to
modify the naive Poisson structure only at these end points by locality. The
most general form of the modified Poisson brackets read
\begin{align}
& \left\{ X^{i}(\sigma),X^{j}(\sigma^{\prime})\right\}  \notag \\
& \hspace{5mm}=(\mathcal{A}_{L})^{ij}\delta(\sigma+\sigma^{\prime })+(%
\mathcal{A}_{R})^{ij}\delta(2\pi-\sigma-\sigma^{\prime}),  \label{3.7} \\
& \left\{ X^{i}(\sigma),P_{j}(\sigma^{\prime})\right\} =\delta_{{\ }%
j}^{i}\delta(\sigma-\sigma^{\prime})  \notag \\
& \hspace{5mm}+(\mathcal{B}_{L})_{{\ }j}^{i}\delta(\sigma+\sigma^{\prime })+(%
\mathcal{B}_{R})_{{\ }j}^{i}\,\delta(2\pi-\sigma-\sigma^{\prime }),
\label{3.8} \\
& \left\{ P_{i}(\sigma),P_{j}(\sigma^{\prime})\right\} =(\mathcal{C}%
_{L})_{ij}\delta(\sigma+\sigma^{\prime})+(\mathcal{C}_{R})_{ij}\delta
(2\pi-\sigma-\sigma^{\prime}),   \label{3.9}
\end{align}
where $\mathcal{A}_{L,R}$, $\mathcal{B}_{L,R}$ and $\mathcal{C}_{L,R}$ are
assumed to be some operators which may act on the variable $\sigma^{\prime}$%
, and $\mathcal{A}_{L,R}$, $\mathcal{C}_{L,R}$ are antisymmetric under $%
i\leftrightarrow j$. The first term on the right hand side of (\ref{3.8})
has to be there since we need to keep the bulk equations of motion
unchanged. Our main task will be the determination of the operators $%
\mathcal{A}_{L,R}$, $\mathcal{B}_{L,R}$, $\mathcal{C}_{L,R}$, so that the
equations (\ref{3.7})-(\ref{3.9}) defines a consistent Poisson structure for
the world sheet theory of the open string.

To determine the values of the operators $\mathcal{A}_{L,R}$, $\mathcal{B}%
_{L,R}$, $\mathcal{C}_{L,R}$, we now reformulate the boundary conditions (%
\ref{3.3}) into the following boundary constraints, in which the world sheet
time derivatives $\partial _{\tau }X^{i}$ are expressed in terms of $P_{i}$
and $X^{j}$ by use of (\ref{3.4}):
\begin{align}
& (G_{L})^{i}\equiv \lim_{\epsilon \rightarrow 0^{+}}\int_{0}^{\pi }d\sigma
\delta (\sigma -\epsilon )[\partial _{\sigma }X^{i}  \notag \\
& \hspace{5mm}+(2\pi \alpha ^{\prime })^{2}B^{im}(X)(P_{m}-B_{mk}(X)\partial
_{\sigma }X^{k})]\simeq 0,  \label{3.10} \\
& (G_{R})^{i}\equiv \lim_{\epsilon \rightarrow 0^{+}}\int_{0}^{\pi }d\sigma
\delta (\pi -\epsilon -\sigma )[\partial _{\sigma }X^{i}  \notag \\
& \hspace{5mm}+(2\pi \alpha ^{\prime })^{2}B^{im}(X)(P_{m}-B_{mk}(X)\partial
_{\sigma }X^{k})]\simeq 0.  \label{3.11}
\end{align}%
One should notice that the order of limitation and integration in the last
two equations cannot be reverted in order that the boundary constraints $%
(G_{L})^{i}$ and $(G_{R})^{i}$ are actually equivalent to the original
boundary conditions.

Through some straightforward calculations with the help of (\ref{3.7})-(\ref%
{3.9}), we can obtain the following Poisson brackets,
\begin{align}
& \{(G_{L})^{i},X^{j}(\sigma ^{\prime })\}=\lim_{\epsilon \rightarrow
0^{+}}\int_{0}^{\pi }d\sigma \delta (\sigma -\epsilon )  \notag \\
& \hspace{6mm}\times \left[ \delta _{{\ }n}^{i}\partial _{\sigma }+(2\pi
\alpha ^{\prime })^{2}\left( \frac{\delta B^{im}}{\delta X^{n}}%
P_{m}-B^{im}B_{mk}\delta _{{\ }n}^{k}\partial _{\sigma }\right. \right.
\notag \\
& \hspace{6mm}\left. \left. -\frac{\delta (B^{im}B_{mk})}{\delta X^{n}}%
\partial _{\sigma }X^{k}\right) \right] \left\{ X^{n}(\sigma ),X^{j}(\sigma
^{\prime })\right\}   \notag \\
& \hspace{6mm}+\lim_{\epsilon \rightarrow 0^{+}}\int_{0}^{\pi }d\sigma
\delta (\sigma -\epsilon )(2\pi \alpha ^{\prime })^{2}B^{im}\left\{
P_{m}(\sigma ),X^{j}(\sigma ^{\prime })\right\}   \notag \\
& \hspace{3mm}=\left[ (2\pi \alpha ^{\prime })^{2}(\mathcal{P}-\mathcal{X})%
\mathcal{A}_{L}+\left( I-(2\pi \alpha ^{\prime })^{2}B^{2}\right) \mathcal{A}%
_{L}\partial _{\sigma ^{\prime }}\right.   \notag \\
& \hspace{6mm}-\left. (2\pi \alpha ^{\prime })^{2}B(I+\mathcal{B}_{L})\right]
^{ij}\delta (\sigma ^{\prime }),  \label{3.12}
\end{align}%
where and hereafter we will always use the abbreviations
\begin{equation*}
\mathcal{P}_{{\ }n}^{i}\equiv \frac{\delta B^{im}}{\delta X^{n}}P_{m},\;\;%
\mathcal{X}_{{\ }n}^{i}\equiv \frac{\delta (B^{2})_{{\ }k}^{i}}{\delta X^{n}}%
\partial _{\sigma }X^{k}.
\end{equation*}%
Analogously, we have
\begin{align}
& \{(G_{L})^{i},P_{j}(\sigma ^{\prime })\}=\lim_{\epsilon \rightarrow
0^{+}}\int_{0}^{\pi }d\sigma \delta (\sigma -\epsilon )  \notag \\
& \hspace{6mm}\times \left[ (2\pi \alpha ^{\prime })^{2}\left( \frac{\delta
B^{im}}{\delta X^{n}}P_{m}-B^{im}B_{mk}\delta _{{\ }n}^{k}\partial _{\sigma
}\right. \right.   \notag \\
& \hspace{6mm}\left. \left. -\frac{\delta (B^{im}B_{mk})}{\delta X^{n}}%
\partial _{\sigma }X^{k}\right) +\delta _{{\ }n}^{i}\partial _{\sigma }%
\right] \{X^{n}(\sigma ),P_{j}(\sigma ^{\prime })\}  \notag \\
& \hspace{6mm}+\lim_{\epsilon \rightarrow 0^{+}}\int_{0}^{\pi }d\sigma
\delta (\sigma -\epsilon )(2\pi \alpha ^{\prime })^{2}B^{im}\{P_{m}(\sigma
),P_{j}(\sigma ^{\prime })\}  \notag \\
& \hspace{3mm}=\{\left[ (2\pi \alpha ^{\prime })^{2}B^{2}-I\right] (I-%
\mathcal{B}_{L})\partial _{\sigma ^{\prime }}  \notag \\
& \hspace{6mm}+(2\pi \alpha ^{\prime })^{2}\left[ (\mathcal{P}-\mathcal{X}%
)(I+\mathcal{B}_{L})+B\mathcal{C}_{L}\right] \}_{{\ }j}^{i}\delta (\sigma
^{\prime }).  \label{3.13}
\end{align}%
With the replacements $\mathcal{A}_{L}\rightarrow \mathcal{A}_{R}$, $%
\mathcal{B}_{L}\rightarrow \mathcal{B}_{R}$, $\mathcal{C}_{L}\rightarrow
\mathcal{C}_{R}$ and $\delta (\sigma ^{\prime }-\epsilon )\rightarrow \delta
(\pi -\epsilon -\sigma ^{\prime })$ in (\ref{3.12})-(\ref{3.13}), we can get
the similar Poisson brackets for $G_{R}$.

In order that the new Poisson brackets (\ref{3.7})-(\ref{3.9}) be consistent
with the boundary conditions (\ref{3.3}), the Poisson brackets (\ref{3.12})
and (\ref{3.13}) have to vanish. This leads to the following operator
equations
\begin{align}
& (2\pi\alpha^{\prime})^{2}(\mathcal{P}-\mathcal{X})\mathcal{A}_{L,R}+\left(
I-(2\pi\alpha^{\prime})^{2}B^{2}\right) \mathcal{A}_{L,R}\partial
_{\sigma^{\prime}}  \notag \\
& \quad-(2\pi\alpha^{\prime})^{2}B(I+\mathcal{B}_{L,R})=0,  \label{3.14} \\
& \left[ (2\pi\alpha^{\prime})^{2}B^{2}-I\right] (I-\mathcal{B}_{L,R})
\partial_{\sigma^{\prime}}  \notag \\
& \quad+(2\pi\alpha^{\prime})^{2}\left[ (\mathcal{P}-\mathcal{X})(I+\mathcal{%
B}_{L,R})+B\mathcal{C}_{L,R}\right] =0,   \label{3.15}
\end{align}
where the action of the operators are right-associative. In particular, $%
\partial_{\sigma^{\prime}}$ acts not only on the operators next to it, but
also on any other quantities to which the left hand side of the complete
equation may be applied. Once the last two equations (\ref{3.14})-(\ref{3.15}%
) are satisfied, the boundary constraints $G_{L,R}$ will Poisson-commute
with everything in the phase space and hence they can be set equal to zero
without causing any inconsistency.

However, contrary to the case with constant background $B$-field, we cannot
finish the story upon getting the two equations (\ref{3.14})-(\ref{3.15}).
We must also require that the modified Poisson brackets (\ref{3.7})-(\ref%
{3.9}) satisfy Jacobi identity. As mentioned just now, since the boundary
constraints $G_{L,R}$ Poisson-commute with everything in the phase space
once (\ref{3.14})-(\ref{3.15}) are satisfied, we cannot expect to get
anything new with $G_{L,R}$ inserted into Jacobi identities. But there are
other instances of Jacobi identities which need a check. In fact, there are
totally 4 instances to be checked: the Jacobi identities for $%
\{X^{i},X^{j},P_{k}\},\{X^{i},P_{j},P_{k}\},\{X^{i},X^{j},X^{k}\}$ and for $%
\{P_{i},P_{j},P_{k}\}$. The first two of these read%
\begin{align}
& \{X^{i}(\sigma),\{X^{j}(\sigma^{\prime}),P_{k}(\sigma^{\prime\prime
})\}\}+\{X^{j}(\sigma^{\prime}),\{P_{k}(\sigma^{\prime\prime}),X^{i}(\sigma)%
\}\}  \notag \\
& \quad+\{P_{k}(\sigma^{\prime\prime}),\{X^{i}(\sigma),X^{j}(\sigma^{\prime
})\}\}=0,  \label{17} \\
& \{X^{i}(\sigma),\{P_{j}(\sigma^{\prime}),P_{k}(\sigma^{\prime\prime
})\}\}+\{P_{j}(\sigma^{\prime}),\{P_{k}(\sigma^{\prime\prime}),X^{i}(\sigma)%
\}\}  \notag \\
& \quad+\{P_{k}(\sigma^{\prime\prime}),\{X^{i}(\sigma),P_{j}(\sigma^{\prime
})\}\}=0.   \label{18}
\end{align}
Straightforward calculations using (\ref{3.7})-(\ref{3.9}) shows that (\ref%
{17}) and (\ref{18}) are equivalent to the following equations at the left
end of the open string,%
\begin{align}
& (\mathcal{A}_{L})^{im}\delta(\sigma+\sigma^{\prime\prime})\frac {%
\delta\lbrack(\mathcal{B}_{L})^{j}{}_{k}\delta(\sigma^{\prime}+\sigma
^{\prime\prime})]}{\delta X^{m}(\sigma^{\prime\prime})}  \notag \\
& +[\delta^{i}{}_{m}\delta(\sigma-\sigma^{\prime\prime})+(\mathcal{B}%
_{L})^{i}{}_{m}\delta(\sigma+\sigma^{\prime\prime})]\frac{\delta \lbrack(%
\mathcal{B}_{L})^{j}{}_{k}\delta(\sigma^{\prime}+\sigma^{\prime\prime })]}{%
\delta P_{m}(\sigma^{\prime\prime})}  \notag \\
& -(\mathcal{A}_{L})^{jm}\delta(\sigma^{\prime}+\sigma^{\prime\prime})\frac{%
\delta\lbrack(\mathcal{B}_{L})_{k}{}^{i}\delta(\sigma+\sigma
^{\prime\prime})]}{\delta X^{m}(\sigma^{\prime\prime})}  \notag \\
& -[\delta^{j}{}_{m}\delta(\sigma^{\prime}-\sigma^{\prime\prime })+(\mathcal{%
B}_{L})^{j}{}_{m}\delta(\sigma^{\prime}+\sigma^{\prime\prime })]\frac{%
\delta\lbrack(\mathcal{B}_{L})_{k}{}^{i}\delta(\sigma+\sigma
^{\prime\prime})]}{\delta P_{m}(\sigma^{\prime\prime})}  \notag \\
& -[\delta_{k}{}^{m}\delta(\sigma^{\prime}-\sigma^{\prime\prime })+(\mathcal{%
B}_{L})_{k}{}^{m}\delta(\sigma^{\prime}+\sigma^{\prime\prime })]\frac{%
\delta\lbrack(\mathcal{A}_{L})^{ij}\delta(\sigma+\sigma^{\prime})]}{\delta
X^{m}(\sigma^{\prime})}  \notag \\
& +(\mathcal{C}_{L})_{km}\delta(\sigma^{\prime\prime}+\sigma^{\prime})\frac{%
\delta\lbrack(\mathcal{A}_{L})^{ij}\delta(\sigma+\sigma^{\prime})]}{\delta
P_{m}(\sigma^{\prime})}=0,   \label{25}
\end{align}%
\begin{align}
& (\mathcal{A}_{L})^{im}\delta(\sigma+\sigma^{\prime\prime})\frac {%
\delta\lbrack(\mathcal{C}_{L})_{jk}\delta(\sigma^{\prime}+\sigma
^{\prime\prime})]}{\delta X^{m}(\sigma^{\prime\prime})}  \notag \\
& -(\mathcal{C}_{L})_{jm}\delta(\sigma^{\prime}+\sigma^{\prime\prime})\frac{%
\delta\lbrack(\mathcal{B}_{L})_{k}{}^{i}\delta(\sigma+\sigma
^{\prime\prime})]}{\delta P_{m}(\sigma^{\prime\prime})}  \notag \\
& +[\delta^{i}{}_{m}\delta(\sigma-\sigma^{\prime\prime})+(\mathcal{B}%
_{L})^{i}{}_{m}\delta(\sigma+\sigma^{\prime\prime})]\frac{\delta \lbrack(%
\mathcal{C}_{L})_{jk}\delta(\sigma^{\prime}+\sigma^{\prime\prime})]}{\delta
P_{m}(\sigma^{\prime\prime})}  \notag \\
& +[\delta_{j}{}^{m}\delta(\sigma^{\prime\prime}-\sigma^{\prime })+(\mathcal{%
B}_{L})_{j}{}^{m}\delta(\sigma^{\prime\prime}-\sigma^{\prime })]\frac{%
\delta\lbrack(\mathcal{B}_{L})_{k}{}^{i}\delta(\sigma+\sigma
^{\prime\prime})]}{\delta X^{m}(\sigma^{\prime\prime})}  \notag \\
& -[\delta_{k}{}^{m}\delta(\sigma^{\prime}-\sigma^{\prime\prime })+(\mathcal{%
B}_{L})_{k}{}^{m}\delta(\sigma^{\prime}+\sigma^{\prime\prime })]\frac{%
\delta\lbrack(\mathcal{B}_{L})^{i}{}_{j}\delta(\sigma+\sigma^{\prime })]}{%
\delta X^{m}(\sigma^{\prime})}  \notag \\
& +(\mathcal{C}_{L})_{km}\delta(\sigma^{\prime\prime}+\sigma^{\prime})\frac{%
\delta\lbrack(\mathcal{B}_{L})^{i}{}_{j}\delta(\sigma+\sigma^{\prime })]}{%
\delta P_{m}(\sigma^{\prime})}=0,   \label{26}
\end{align}
and similar ones at the right end of the open string with $\mathcal{A}_{L},%
\mathcal{B}_{L},\mathcal{C}_{L}\leftrightarrow\mathcal{A}_{R},\mathcal{B}%
_{R},\mathcal{C}_{R}$, and $\sigma,\sigma^{\prime},\sigma
^{\prime\prime}\leftrightarrow\pi-\sigma,\pi-\sigma^{\prime},\pi
-\sigma^{\prime\prime}$ in all the $\delta$-functions.

Using the fact that
\begin{equation*}
\frac{\delta}{\delta X^{m}(\sigma^{\prime\prime})}\delta(\sigma+\sigma
^{\prime\prime})=\frac{\delta}{\delta P_{m}(\sigma^{\prime\prime})}%
\delta(\sigma+\sigma^{\prime\prime})=0
\end{equation*}
and that $\delta(\sigma+\sigma^{\prime\prime})\delta(\sigma^{\prime}+%
\sigma^{\prime\prime})=\delta(\sigma+\sigma^{\prime\prime})\delta
(\sigma^{\prime}-\sigma^{\prime\prime})$ is nonzero only at $\sigma
=\sigma^{\prime}=\sigma^{\prime\prime}=0$ if $\sigma,\sigma^{\prime},%
\sigma^{\prime\prime}$ are all non-negative, we may drop all the $\delta $%
-function dependencies in (\ref{25}) and (\ref{26}) and get (now the
equations at both ends of the open string can be written in a unified way)%
\begin{align}
& (\mathcal{A}_{L,R})^{im}\frac{\delta\lbrack(\mathcal{B}_{L,R})^{j}{}_{k}]}{%
\delta X^{m}}+[\delta^{i}{}_{m}+(\mathcal{B}_{L,R})^{i}{}_{m}]\frac {%
\delta\lbrack(\mathcal{B}_{L,R})^{j}{}_{k}]}{\delta P_{m}}  \notag \\
& \quad-(\mathcal{A}_{L,R})^{jm}\frac{\delta\lbrack(\mathcal{B}%
_{L,R})_{k}{}^{i}]}{\delta X^{m}}-[\delta^{j}{}_{m}+(\mathcal{B}%
_{L,R})^{j}{}_{m}]\frac{\delta\lbrack(\mathcal{B}_{L,R})_{k}{}^{i}]}{\delta
P_{m}}  \notag \\
& \quad-[\delta^{m}{}_{k}+(\mathcal{B}_{L,R})^{m}{}_{k}]\frac{\delta \lbrack(%
\mathcal{A}_{L,R})^{ij}]}{\delta X^{m}}+(\mathcal{C}_{L,R})_{km}\frac{%
\delta\lbrack(\mathcal{A}_{L,R})^{ij}]}{\delta P_{m}}  \notag \\
& =0,   \label{27}
\end{align}%
\begin{align}
& (\mathcal{A}_{L,R})^{im}\frac{\delta\lbrack(\mathcal{C}_{L,R})_{jk}]}{%
\delta X^{m}}+[\delta^{i}{}_{m}+(\mathcal{B}_{L,R})^{i}{}_{m}]\frac {%
\delta\lbrack(\mathcal{C}_{L,R})_{jk}]}{\delta P_{m}}  \notag \\
& \quad-(\mathcal{C}_{L,R})_{jm}\frac{\delta\lbrack(\mathcal{B}%
_{L,R})_{k}{}^{i}]}{\delta P_{m}}+[\delta_{j}{}^{m}+(\mathcal{B}%
_{L,R})_{j}{}^{m}]\frac{\delta\lbrack(\mathcal{B}_{L,R})_{k}{}^{i}]}{\delta
X^{m}}  \notag \\
& \quad-[\delta_{k}{}^{m}+(\mathcal{B}_{L,R})_{k}{}^{m}]\frac{\delta \lbrack(%
\mathcal{B}_{L,R})^{i}{}_{j}]}{\delta X^{m}}+(\mathcal{C}_{L,R})_{km}\frac{%
\delta\lbrack(\mathcal{B}_{L,R})^{i}{}_{j}]}{\delta P_{m}}  \notag \\
& =0.   \label{28}
\end{align}
By similar arguments, we get the following equations from the other two
instances of Jacobi identities involving three $X$'s or three $P$'s, which
are also required to hold at both ends of the open string,%
\begin{align}
& \frac{\delta(\mathcal{A}_{L,R})^{ij}}{\delta X^{m}}(\mathcal{A}%
_{L,R}){}^{mk}-\frac{\delta(\mathcal{A}_{L,R})^{ij}}{\delta P_{m}}\left[
\delta _{m}{}^{k}+(\mathcal{B}_{L,R})_{m}{}^{k}\right]  \notag \\
& \quad+\frac{\delta(\mathcal{A}_{L,R})^{jk}}{\delta X^{m}}(\mathcal{A}%
_{L,R}){}^{mi}-\frac{\delta(\mathcal{A}_{L,R})^{jk}}{\delta P_{m}}\left[
\delta_{m}{}^{i}+(\mathcal{B}_{L,R})_{m}{}^{i}\right]  \notag \\
& \quad+\frac{\delta(\mathcal{A}_{L,R})^{ki}}{\delta X^{m}}(\mathcal{A}%
_{L,R}){}^{mj}-\frac{\delta(\mathcal{A}_{L,R})^{ki}}{\delta P_{m}}\left[
\delta_{m}{}^{j}+(\mathcal{B}_{L,R})_{m}{}^{j}\right]  \notag \\
& \,=0,   \label{29}
\end{align}%
\begin{align}
& \frac{\delta(\mathcal{C}_{L,R})_{ij}}{\delta X^{m}}\left[
\delta^{m}{}_{k}+(\mathcal{B}_{L,R})^{m}{}_{k}\right] +\frac{\delta(\mathcal{%
C}_{L,R})_{ij}}{\delta P_{m}}(\mathcal{C}_{L,R})_{mk}  \notag \\
& \quad+\frac{\delta(\mathcal{C}_{L,R})_{jk}}{\delta X^{m}}\left[ \delta
^{m}{}_{i}+(\mathcal{B}_{L,R})^{m}{}_{i}\right] +\frac{\delta(\mathcal{C}%
_{L,R})_{jk}}{\delta P_{m}}(\mathcal{C}_{L,R})_{mi}  \notag \\
& \quad+\frac{\delta(\mathcal{C}_{L,R})_{ki}}{\delta X^{m}}\left[ \delta
^{m}{}_{j}+(\mathcal{B}_{L,R})^{m}{}_{j}\right] +\frac{\delta(\mathcal{C}%
_{L,R})_{ki}}{\delta P_{m}}(\mathcal{C}_{L,R})_{mj}  \notag \\
& \,=0.   \label{30}
\end{align}

The set of equations (\ref{3.14})-(\ref{3.15}) and (\ref{27})-(\ref{30})
form the complete set of equations which the operators $\mathcal{A}_{L,R},%
\mathcal{B}_{L,R},\mathcal{C}_{L,R}$ must obey. However, this is a system of
$6$ equations for $3$ unknowns, i.e. they are over-determined and a general
solution is not guaranteed to exist. In particular, although we do not have
a rigorous proof for the non-existence of solutions with nonzero $\mathcal{A}%
_{L,R}$, it is extremely difficult to find one. On the contrary, it is very
easy to see that the case $\left( \mathcal{A}_{L,R}\right) ^{ij}=0$ is a
solution, with%
\begin{align*}
& \left( \mathcal{B}_{L,R}\right) ^{i}{}_{j}=-\delta^{i}{}_{j}, \\
& \left( \mathcal{C}_{L,R}\right) _{ij}=\frac{2}{(2\pi\alpha^{\prime})^{2}}
\\
& \quad\times\left( G+2\pi\alpha^{\prime}B\right) _{ik}(B^{-1})^{kl}\left(
G-2\pi\alpha^{\prime}B\right) _{lj}\partial_{\sigma^{\prime}}.
\end{align*}
Inserting the above solution for $\mathcal{A}_{L,R},\mathcal{B}_{L,R},%
\mathcal{C}_{L,R}$ into the earlier postulation (\ref{3.7})-(\ref{3.9}) we
get the following consistent Poisson structure,%
\begin{align}
& \left\{ X^{i}(\sigma),X^{j}(\sigma^{\prime})\right\} =0,  \label{poi1} \\
& \left\{ X^{i}(\sigma),P_{j}(\sigma^{\prime})\right\}  \notag \\
& \quad=\delta_{{\ }j}^{i}\left[ \delta(\sigma-\sigma^{\prime})-\delta(%
\sigma+\sigma^{\prime})-\delta(2\pi-\sigma-\sigma^{\prime})\right] ,
\label{poi2} \\
& \left\{ P_{i}(\sigma),P_{j}(\sigma^{\prime})\right\} =\frac{2}{%
(2\pi\alpha^{\prime})^{2}}  \notag \\
& \quad\times\left( G+2\pi\alpha^{\prime}B\right) _{ik}(B^{-1})^{kl}\left(
G-2\pi\alpha^{\prime}B\right) _{lj}  \notag \\
& \quad\times\partial_{\sigma^{\prime}}\left[ \delta(\sigma+\sigma^{\prime
})+\delta(2\pi-\sigma-\sigma^{\prime})\right] .   \label{poi3}
\end{align}

It is remarkable that the form of the above Poisson structure coincides with
the $\left( \mathcal{A}_{L,R}\right) ^{ij}=0$ solution in our previous paper
\cite{He}, though the background fields $G_{ij}$ and $B_{ij}$ are now both
varying with spacetime. Therefore, we may call the Poisson brackets (\ref%
{poi1})-(\ref{poi3}) the \textquotedblleft background
independent\textquotedblright\ Poisson structure for the open string theory.
It is also worth mentioning that the Poisson brackets (\ref{poi1})-(\ref%
{poi3}) does not depend on the detailed choice of the background fields,
therefore they are also valid for general nonlinear sigma models defined by (%
\ref{3.1}) with the fields $G_{ij}(X)$ and $B_{ij}(X)$ not necessarily
preserving conformal symmetry. Let us remark that the Poisson structure (\ref%
{poi1})-(\ref{poi3}) is an \emph{exact} Poisson structure for any background
fields $G_{ij}$ and $B_{ij}$, unlike the approximate result of \cite{Ho1},
which works only for a non-oscillating string or a rigid rod.

\section{Discussions on quantization}

Having obtained a consistent set of Poisson brackets (\ref{poi1})-(\ref{poi3}%
) for the open string theory, we now come to the step for quantization. The
Poisson brackets (\ref{poi1})-(\ref{poi2}) are both linear and can be
changed into canonical equal time commutators via the substitution $%
\{\,,\,\}\rightarrow -i[\,,\,]$, i.e.
\begin{align}
& \left[ X^{i}(\sigma ),X^{j}(\sigma ^{\prime })\right] =0,  \label{com1} \\
& \left[ X^{i}(\sigma ),P_{j}(\sigma ^{\prime })\right]   \notag \\
& \quad =i\delta _{{\ }j}^{i}\left[ \delta (\sigma -\sigma ^{\prime
})-\delta (\sigma +\sigma ^{\prime })-\delta (2\pi -\sigma -\sigma ^{\prime
})\right] .  \label{com2}
\end{align}%
The last one of the Poisson brackets, eq. (\ref{poi3}), is in general
nonlinear in the spacetime coordinates $X^{i}$, therefore to change this
last Poisson bracket into commutators we need to pay some more care on the
operator ordering on the right hand side. Fortunately, since the right hand
side of (\ref{poi3}) depends only on $X^{i}$, and we have seen from (\ref%
{com1}) that the operators $X^{i}$ commute among themselves at equal world
sheet time, the operator ordering problem on the right hand side of (\ref%
{poi3}) can be easily resolved. The quantized form of (\ref{poi3}) is then
given as%
\begin{align}
& \left[ P_{i}(\sigma ),P_{j}(\sigma ^{\prime })\right] =i\frac{2}{(2\pi
\alpha ^{\prime })^{2}}  \notag \\
& \quad \times :\left( G+2\pi \alpha ^{\prime }B\right)
_{ik}(B^{-1})^{kl}\left( G-2\pi \alpha ^{\prime }B\right) _{lj}:  \notag \\
& \quad \times \partial _{\sigma ^{\prime }}\left[ \delta (\sigma +\sigma
^{\prime })+\delta (2\pi -\sigma -\sigma ^{\prime })\right] .  \label{com3}
\end{align}%
The equations (\ref{com1})-(\ref{com3}) constitute the set of fundamental
commutators for the quantized open string theory in generic background. We
can see that the coordinates at the ends of the strings are free of
noncommutativity, thanks to (\ref{com1}). Moreover, the commutators between
the coordinates and momenta \emph{at the boundaries} are also vanishing (see
eq.(\ref{com2}) at $\sigma =\sigma ^{\prime }=0,\pi $). The only
noncommutativity to appear in the quantum theory is in between the momenta,
due to (\ref{com3}). It then follows from the standard Heisenberg equations
that the world sheet time derivatives $\partial _{\tau }X^{i}$ at $\sigma
=0,\pi $ will be identically zero, implying that the branes to which the
string ends are attached can only be D-branes, even if classically the
boundary conditions appears to be mixed. Though we cannot say that the above
quantization scheme is the only possible one for open string in generic
background, we can at least conclude that the \emph{spacetime}
noncommutativity at the boundary D-branes \emph{can be avoided}, and should
be avoided, since noncommutative boundaries would in general spoil the
stability of the D-branes, i.e. $\partial _{\tau }X^{i}$ can become nonzero
at $\sigma =0,\pi $. That spacial noncommutative boundaries could be avoided
for open string in generic background might also be helpful in settling the
embarrassing ambiguities in the Seiberg-Witten maps which are so far
unresolved \cite{AMB}.

Before ending this Letter, we should mention that we are not
attempting to make a systematic description for the quantum theory
of open strings in a generic background. For that purpose we
should have gone done into details on the treatment of Virasoro
constraints in the bulk (which is a consequence of the bulk
equation of motion for the world sheet metric) as well as the
analysis on the physical spectrum. However, since the Virasoro
constraints are constraints purely in the bulk because on the
boundaries the variations of the world sheet metric should simply
be vanishing, giving rise to no constraints over the world sheet
fields at all, our results on the boundary commutators should not
be affected by the bulk Virasoro constraints. On the other hand,
the analysis of physical spectrum in the presence of generic
background is a very difficult task if not impossible, because
such an analysis depends crucially on the mode expansions for the
solutions of bulk fields, the form of which is not known unless
the explicit form of the background fields is given. We therefore
simply skip these steps by stressing once again that the boundary
commutators between the world sheet fields will not be affected by
such analysis.

After the completion of this work, we are informed by the authors of \cite%
{Subir} and \cite{Schiappa} about their works. These works also
considered the problem of quantization for open string in curved
metric and non-constant $B$-fields, and both of them give
noncommutive results. However, the validity of Jacobi identities
are not checked in \cite{Subir}, while in \cite{Schiappa}, it is
explicitly stated that the results are based on the violation of
Jacobi identities. Our results in this Letter is on the complete
contrary: we showed that the spacial noncommutativity on the
string ends
does not necessarily appear, and that when the string ends are \emph{%
commutative}, there is no violation of Jacobi identities.

%\vspace{3mm}

\section*{Acknowledgement}

This work is supported in part by the National Natural Science Foundation of
China.

\end{document}